\documentclass[a4paper,11pt]{article}
\usepackage[utf8]{inputenc}
\usepackage{authblk}
\usepackage{graphicx}
\usepackage{amsmath}
\usepackage[hidelinks]{hyperref}
\usepackage{threeparttable}
\usepackage{comment}

\usepackage{natbib}
\setcitestyle{authoryear,open={(},close={)}}

\usepackage[english,bottom,portrait,draft]{draftcopy} 

\usepackage[margin=1in,footskip=0.25in]{geometry}
\usepackage{array}
\usepackage{arydshln}
\usepackage{xcolor}

\newcolumntype{P}[1]{>{\centering\arraybackslash}p{#1}}
\newcolumntype{M}[1]{>{\centering\arraybackslash}m{#1}}

\title{Modeling and forecasting the early evolution of the Covid-19 pandemic in Brazil}

\author[1]{Saulo B. Bastos }

\author[1,2,3]{Daniel O. Cajueiro}

\affil[1]{Departamento de Economia, FACE, Universidade de Bras\'{i}lia (UnB), Campus Universit\'{a}rio Darcy Ribeiro, 70910-900, Bras\'{i}lia, Brazil.}

\affil[2]{Nacional Institute of Science and Technology for Complex Systems (INCT-SC).}

\affil[3]{LAMFO, FACE - Universidade de Bras\'{i}lia (UnB), Campus Universit\'{a}rio Darcy Ribeiro, 70910-900, Bras\'{i}lia, Brazil.}

\date{\today}
\begin{document}

\maketitle

\begin{abstract}
We model and forecast the early evolution of the COVID-19 pandemic in Brazil using Brazilian recent data from February 25, 2020 to March 30, 2020. This early period accounts for unawareness of the epidemiological characteristics of the disease in a new territory, sub-notification of the real numbers of infected people and the timely introduction of social distancing policies to flatten the spread of the disease. We use two variations of the SIR model and we include a parameter that comprises the effects of social distancing measures.  Short and long term forecasts show that the social distancing policy imposed by the government is able to flatten the pattern of infection of the COVID-19. However, our results also show that if this policy does not last enough time, it is only able to shift the peak of infection into the future keeping the value of the peak in almost the same value. Furthermore, our long term simulations forecast the optimal date to end the policy. Finally, we show that the proportion of asymptomatic individuals affects the amplitude of the peak of symptomatic infected, suggesting that it is important to test the population.
\end{abstract}

\section*{Introduction}

The world has seen an ongoing pandemic of COVID-19 (coronavirus 2) caused by severe acute respiratory syndrome SARS-CoV-2. According to the World Health Organization (WHO) \citep{who20200329}, although most people infected with it will present mild respiratory symptoms, or no signs of the disease, and recover without needing special treatment, older people, and those with severe medical conditions like diabetes, cardiovascular disease, or chronic respiratory disease may develop serious illness. While the COVID-19 outbreak was first identified  in Wuhan, Hubei, China, in December 2019, we could only confirm the first case in Brazil on February 25, 2020. The first known patient in Brasil was a 61-year-old man from São Paulo who had returned from Lombardy (Italy) and tested positive for the virus. Since then, we may confirm 4579 cases and 159 deaths (March 30, 2020) in roughly the entire Brazilian territory. Like in the rest of the world  \citep{Adam2020}, the Brazilian government response to the pandemic has been the introduction of measures to ensure social distancing, such as schools closure, restricting commerce, banning public events and home office.

We use the Brazilian recent data from February 25, 2020 to March 30, 2020 to model and forecast the evolution of the COVID-19 pandemic. Our study focuses on the early period of the pandemics that accounts for unawareness of the epidemiological characteristics of the disease in a new territory, sub-notification of the real numbers of infected people and the timely introduction of social distancing policies to flatten the spread of the disease. This work has had the practical appeal for providing preliminary estimates of Covid-19 epidemiological parameters and the duration of the  social distancing policy in Brazil. 

The computational modeling of infectious diseases comprises a large collection of models  \citep{Grassly:2008,keeling2011,Brauer2019}. In order to model the evolution of the Covid-19 in Brazil we modify two versions of the the Susceptible-Infected-Recovered (SIR) model  \citep{Kermack1927} to consider the effects of social distancing measures in the evolution of the disease.  The SIR model describes the spread of a disease in a population split into three non-intersecting classes: Susceptible (S)  are individuals who are healthy but can contract the disease; Infected (I) are individuals who are sick; Recovered (I) are individuals who recovered from the disease. Due to the evolution of the disease, the size of each of these classes change over time and the total population size \(N\) is the sum of these classes

\begin{equation}
N(t)=S(t)+I(t)+R(t)\label{eq:Nconstant}. 
\end{equation}

Let \(\beta\) be the average number of contacts that are sufficient for transmission of a person per unit of time \(t\). Then \(\beta I/N\) is the average number of contacts that are sufficient for transmission with infective individuals per unit of time of one susceptible and \((\beta I/N)S\) is the number of new cases per unit of time due to the \(S\) susceptible individuals. Furthermore, let \(\gamma\) be the recovery rate, which is the rate that infected individuals recover or die, leaving the infected class, at constant per capita probability per unit of time.

Based on these definitions, we can write the SIR model as

\begin{equation}
    \begin{array}{rcl}
    \displaystyle \frac{dS}{dt} & = & \displaystyle -\frac{\beta IS}{N}\\[3mm]
    \displaystyle \frac{dI}{dt} & = & \displaystyle \frac{\beta IS}{N} - \gamma I \\[3mm]
    \displaystyle \frac{dR}{dt} & = & \displaystyle \gamma I \\
    \end{array}\;\;\;\textrm{\bf [SIR]}. 
\label{eq:SIR}\end{equation}

It is worth mentioning that we can also evaluate the number of recovered individuals from Eq. (\ref{eq:Nconstant}) using also the number of susceptible and infected individuals, since in this version of the SIR model [Eq. (\ref{eq:SIR})] the population is constant. Actually, since we are modeling a short term pandemic, we do not consider the demographic effects and we assume that an individual does not contract the disease twice. We do not implement this model, we only included it for the sake of reference.

We actually want to estimate the fraction of people that die from the disease. Then we include a probability \(\rho\) of an individual in the class \(I\) dying from infection before recovering  \citep{keeling2011}. In this case, we get the following set of equations

\begin{equation}
    \begin{array}{rcl}
    \displaystyle \frac{dS}{dt} & = & \displaystyle - \frac{\beta I S}{N}\\[3mm]
    \displaystyle \frac{dI}{dt} & = & \displaystyle \frac{\beta I S}{N} - \gamma I  - \frac{\rho}{1 - \rho} \gamma I = \displaystyle \frac{\beta I S}{N} - \frac{\gamma I}{1 - \rho}\\[3mm]
    \displaystyle \frac{dR}{dt} & = & \displaystyle \gamma I \\[3mm]
    \displaystyle \frac{dD}{dt} & = & \displaystyle \frac{\rho}{1-\rho} \gamma I\\
    \end{array}\;\;\;\textrm{\bf [SIRD]},   
\label{eq:SIRDead}
\end{equation}
\noindent where \(\frac{\rho}{1-\rho} \gamma I\) is the number of people in the population that die due to the disease per unity of time and \(D\) is the number of people that die due to the disease. Note that in this case the number of individuals in the population reduces due to the infection according to $\frac{dN}{dt} = -\frac{\rho}{1 - \rho} \gamma I$. For the ease of reference, we call this model ``{\bf SIRD}'' (Susceptible-Infected-Recovered-Dead) model.

Since, in the case of the COVID-19, there is a relevant percentage of the infected individuals that are asymptomatic, we split the class of infected individuals in symptomatic and asymptomatic  \citep{Robinson2013,Arino2008,Longini2004}: 


\begin{equation}
\begin{array}{rcl}
\displaystyle \frac{dS}{dt} & = & \displaystyle  - (\beta_A I_A + \beta_S I_S)\frac{ S}{N} \\[3mm]
\displaystyle \frac{dI_A}{dt} & = & \displaystyle  (1-p)(\beta_A I_A + \beta_S I_S)\frac{ S}{N} - (\gamma_A) I_A \\[3mm]
\displaystyle \frac{dI_S}{dt} & = & p (\beta_A I_A + \beta_S I_S)\frac{ S}{N} - \frac{\gamma_S I_S}{1 - \rho} \\[3mm]
\displaystyle \frac{dR_A}{dt} & = & \displaystyle \gamma_A I_A \\[3mm]
\displaystyle \frac{dR_S}{dt} & = & \displaystyle \gamma_S I_S \\[3mm]
\displaystyle \frac{dD}{dt} & = & \displaystyle \frac{\rho}{1 - \rho} \gamma_S I_S \\[3mm]
\end{array}\;\;\;\textrm{{\bf [SIRASD]}},
\label{eq:IAISwithP} 
\end{equation}

\noindent where $I_A$ is the number of asymptomatic individuals, $I_S$ is the number of symptomatic individuals, $R_A$ and $R_S$ are the recovered individuals from the asymptomatic and symptomatic infection, respectively, and $p$ is the proportion of individuals who develop symptoms. For ease of reference, we call this model ``{\bf SIRASD}'' (Susceptible-Infected-Recovered for Asymptomatic-Symptomatic and Dead) model. Like the SIRD model, the condition that $N$ is constant does not hold anymore and if we need to evaluate $N$ over time, we need to integrate $\frac{dN}{dt} = -\frac{\rho}{1 - \rho} \gamma_S I_S$.

In order to consider the effect of the social distancing policy, we modify the transmission factors of Eqs. (\ref{eq:SIRDead}) and (\ref{eq:IAISwithP}) by multiplying them by a parameter $\psi \in [0, 1]$, when the date belongs to the period of the implementation of government policy. Otherwise, we use \(\psi=1\). To be precise, we replace \(\beta\) in Eq. (\ref{eq:SIRDead}) by \(\psi \beta\),  \(\beta_A\) in Eq. (\ref{eq:IAISwithP}) by \(\psi \beta_A\) and \(\beta_S\) in Eq. (\ref{eq:IAISwithP}) by \(\psi \beta_S\).  Note that doing this procedure we avoid the introduction and estimations of new ``\(\beta\)s'' and we may use \(\psi\) to evaluate the effectiveness of social distancing policy. In the end, we may measure the social distance as $1-\psi$. 

Our models provide estimates of the epidemiological parameters, that are consistent with the international literature, and good forecasts of the short-term  Brazilian time series of infected individuals in Brazil. Furthermore, one of our models assesses the number of asymptomatic (or individuals with mild symptoms that do not look for the hospitals and are not being tested). We use these models to simulate long-term scenarios of the pandemics that depend on the level of engagement of the Brazilian social distancing policy. We show that: (1)  The social distancing policy imposed by the government is able to flatten the pattern of contamination provided by the COVID-19; (2) There is an optimal date for abandoning the social distancing policy; (3) Short-term social distancing policies only shift the peak of infection into the future keeping the value of the peak in almost the same value. (4) The proportion of asymptomatic individuals affects the amplitude of the peak of symptomatic infected, meaning that it is important to invest in testing the population, massively or by random sampling.

Our work relates to the recent interesting contributions  \citep{Kucharski2020,Berger2020,Read2020,walker2020} in the sense that all these works try to model the spread of the COVID-19 and to evaluate the countermeasures against this virus. However, our paper differs from these works in the following dimensions: (1) Data: Our work focuses in Brazilian data. This is an important characteristic since different countries may present different demographies and we know that the COVID-19 is riskier for older populations that appear with higher proportion in developed countries. Furthermore, the level of nutrition of the population of the country may affect the probability of contracting and developing the disease. The quality of data may vary from developed countries to underdeveloped ones and, in our paper, we do not use data from other countries to calibrate our models. (2) Model: We use variations of the SIR model mentioned above. One of the advantages of the SIR model is the simplicity and researchers have used this model in several successful attempts to model the spread of infectious diseases  \citep{Shaman2013,Berge2017,Osthus2017,Khaleque2017}.  (3) Estimation: Our paper estimates all the parameters based on a clear hierarchical procedure based on squared error minimization.

\section*{Results}

\subsection*{Data analysis}

\begin{table}
\centering
 \begin{threeparttable}
 \begin{tabular}{M{1.5cm}M{1.5cm}M{3cm}M{9cm}}
 \textbf{Model} & \textbf{Parameter} & \textbf{Value} & \textbf{Other sources} \\\hline
 SIRD& $\beta$ & 0.4417 (0.3695-0.6043) & ---\\ \hline
 SIRD& $\gamma$ & 0.1508 (0.0714-0.3295) & 1/10 to 1/2  \citep{whoSituationReport7} \\ \hline
 SIRD& $\rho$ & 0.0292 (0.0100-0.0485) & 0.049 released by WHO \citep{who20200401} in 2020-04-01, 0.028 in 2020-03-27 released by Brazilian Ministry of Health \citep{BMH20200327}, 0.032 in 2020-03-29 released by Brazilian Ministry of Health \citep{BMH20200329} and 0.014  (0.009-0.021) in Wuhan \citep{Wu2020},\\ \hline
 SIRD & $R_0$ & 2.8421 (1.8142-4.9886) &  3.8 (3.6-4.0) \citep{Read2020} and 2.68 (2.47-2.86) \citep{Wu2020b} in early stages of the disease in China. 2.76 to 3.25 in Italy  \citep{Remuzzi2020}. 2.28 (2.06-2.52) \citep{Zhang2020} for the passengers of the Diamond Princess cruise.\\ \hline
 SIRASD & $\beta_S$ & 0.4417 (0.3695-0.6043) & ---\\ \hline
 SIRASD & $\gamma_S$ & 0.1508 (0.0714-0.3295) & ---\\ \hline
 SIRASD & $R_S$ & 1.1807 (0.5281-1.8613) & ---\\ \hline
 SIRASD & $\beta_A$ & 0.4417 (0.4417-0.4417) & ---\\ \hline
 SIRASD & $\gamma_A$ & 0.1260 (0.1130-0.1445) & ---\\ \hline
 SIRASD & $R_A$ & 2.4209 (1.9143-2.7083) & ---\\ \hline
 SIRASD & $R_0$ & 3.6017 (2.5933-4.1529) & The same as above.\\ \hline 
 SIRASD & $\rho$ & 0.0347 (0.0175-0.0527) & The same as above.\\ \hline
 SIRASD & $p$ & 0.3210 (0.2916-0.3736) & 0.821 (0.798-0.845) for the passengers of the Diamond Princess Cruise  \citep{Mizumoto2020}. 0.692 (0.462-0.923) for the Japanese citizens evacuated from Wuhan \citep{Nishiura2020}.\\ \hline
 \end{tabular} 

 \begin{tablenotes}
      \scriptsize
      \item Notes:
      \item (1) In the SIRD model, $R_0 = \beta(1-\rho)/\gamma$. In the SIRASD model, $R_A = (1-p)\beta_A/\gamma_A$ and $R_S = p \beta_S(1-\rho)/\gamma_S$ and $R_0 = R_A + R_S$.
      \item (2) Some parameters have not presented relevant variation in the significance level of this study. In these cases, the 90\% interval includes only the value of the parameter.
    \end{tablenotes}
 
 \end{threeparttable}

 \caption{Estimated values of the epidemiological parameters.}
\label{tab:epidemiologicalParameters} 
\end{table}

\begin{figure}
    \centering
\begin{tabular}{c}
\includegraphics[width=80mm]{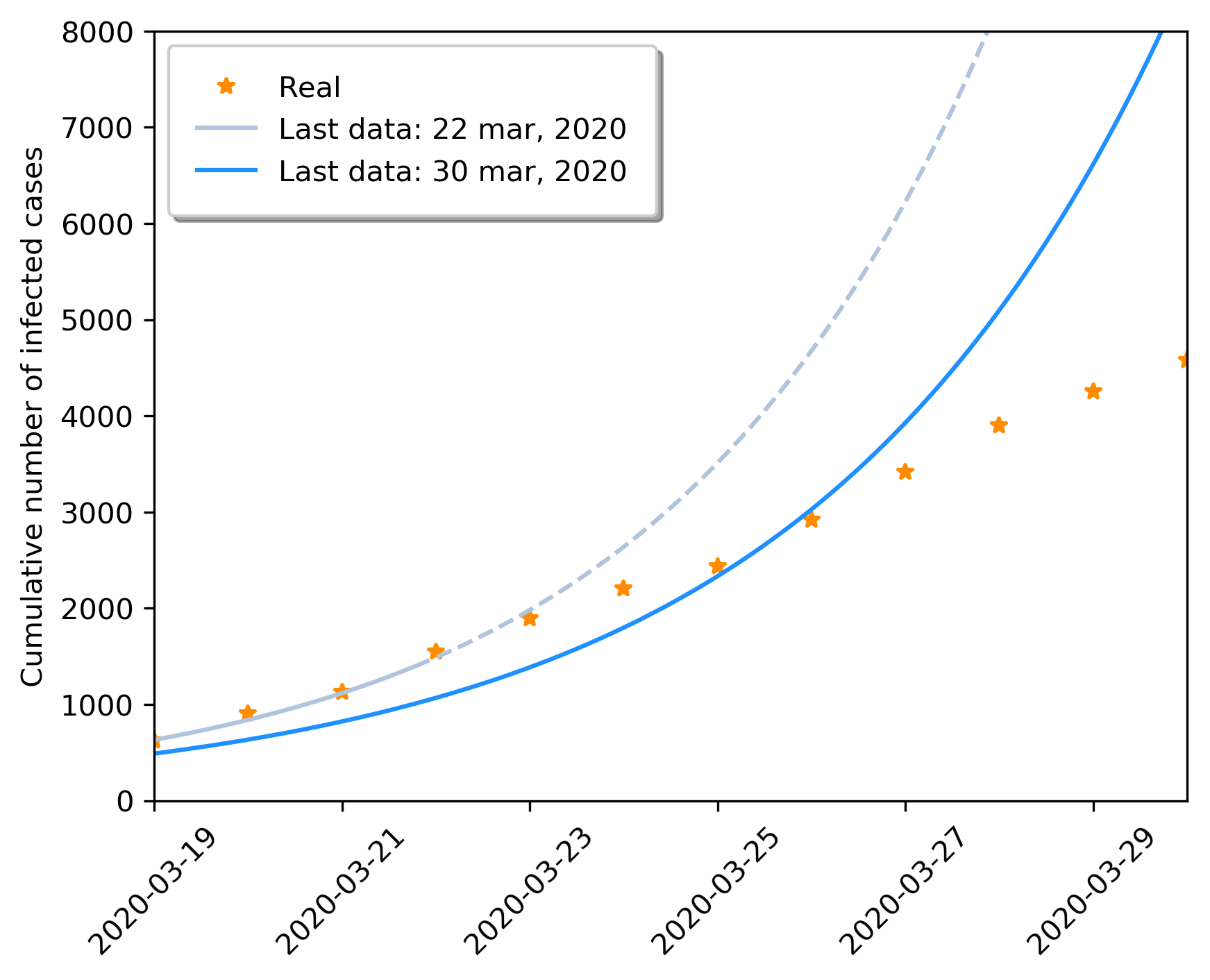}\\
\end{tabular}           
\caption{Estimations of the SIRD model for different final date points. The solid line corresponds to the last date which the model was estimated, and the dashed line are model predictions. We represent the real data as points.}
\label{fig:comparingSocialDistanceSIR}
\end{figure}

\begin{table}
\centering
 \begin{tabular}{ccc@{\extracolsep{10pt}}cc}
 \hline
& \multicolumn{2}{c}{SIRD} & \multicolumn{2}{c}{SIRASD} \\ \cline{2-3} \cline{4-5}
Date & $\psi$ & $R$ & $\psi$ & $R$ \\ \hline
03-23-2020 & 0.8182 & 2.325 & 0.8799 & 2.8968 \\
03-24-2020 & 0.7471 & 2.123 & 0.7786 & 2.5633 \\
03-25-2020 & 0.6639 & 1.887 & 0.6891 & 2.2685 \\
03-26-2020 & 0.6526 & 1.854 & 0.6510 & 2.1433 \\
03-27-2020 & 0.6464 & 1.837 & 0.6409 & 2.1100 \\
03-28-2020 & 0.6421 & 1.825 & 0.6302 & 2.0747 \\
03-29-2020 & 0.6356 & 1.806 & 0.6190 & 2.0379 \\
03-30-2020 & 0.6254 & 1.777 & 0.6156 & 2.0267 \\
\hline
 \end{tabular} 
 \caption{Estimated values of $\psi$ for the SIRD and SIRASD models and the impact on the basic reproductive number $R$.}
\label{tab:psiOverTime} 
\end{table}

We use the real data provided by the Ministry of Health of Brazil from February 25, 2020 to March 30, 2020 in our estimations. If we change the final date of the period of estimation of the epidemiological parameters of the model, we note that there is a structural change in the data suggesting the effectiveness of the social distancing policy. It is worth mentioning that it is hard to know exactly when social distance measures took effect mostly because there is a variable incubation period of the virus given by a range from 2 to 10 days \citep{whoSituationReport7} and some initiatives of social distance measures (such as home office) started even before the official implementation of the social distancing policy. In fact, after March 23, 2020, we are able to see in the data three consecutive reductions in the first difference of the cumulative number of infections, so depending on the final date that is used for the estimation of the SIRD model, the estimated parameters cannot fit the real data anymore, as shown in Figure \ref{fig:comparingSocialDistanceSIR}. Thus, we define two estimation periods: (i) February 25, 2020 to March 22, 2020, in which we estimate the epidemiological parameters of Eqs. (\ref{eq:SIRDead}) and (\ref{eq:IAISwithP}); (ii) March 23, 2020 to March 30, 2020, in which we estimate the paramter \(\psi\).


Regarding the estimation of the epidemiological parameters of Eqs. (\ref{eq:SIRDead}) and (\ref{eq:IAISwithP}), we estimate all parameters of our model by minimizing the squared error of integrated variables and their real values  \citep{Bard1974,Brauer2019}. We proceed in a hierarchical procedure. We start by estimating the parameters of the SIRD model, namely \(\beta\), \(\gamma\) and \(\rho\) by minimizing the squared error  
\begin{equation}
\begin{array}{cc}
\min_{\beta, \gamma, \rho} & \frac{1}{2} \left( \sum_{t}{ f \left( [(I_t^{cum} - D_t) - (\hat{I}_t+\hat{R}_t)]^2 \right) + f\left([D_t - \hat{D}_t]^2\right)} \right)
\end{array},
\label{eq:SIR_model_params_calculation}
\end{equation}
\noindent where $I_t^{cum}$ and $D_t$ are the cumulative number of infected individuals and deaths, which are the real data provided by the Ministry of Health of Brazil, and $\hat{I}_t$, $\hat{R}_t$ and $\hat{D}_t$ are estimated values of the infected, recovered and deaths, respectively. We use the nonlinear function $f(z) = C^2 \log \left( (g(z)/C)^2 \right)$ to correct the exponential characteristic of the series so that the errors of the last values of the series do not dominate the minimization, where \(g(z)=\log(1+z)\). Furthermore, we use the scaling  parameter $C=2$ to soft threshold between inliers and outliers. Using this procedure, we note that the estimated epidemiological parameters vary less among simulations with different random seeds. 

After estimating the SIRD model, we proceed by estimating the SIRASD model. Note that we lack information on the number of asymptomatic individuals, since the clear recommendation of the Ministry of Health is to test for the virus only if one has moderate or severe symptoms. Otherwise, follow the ``stay at home'' policy, which recommends individuals with mild symptoms to stay at home and do not seek for medical attention. Furthermore, the mortality rate is evaluated mostly over the symptomatic ones, since the asymptomatic are in many cases not tested. Therefore, we suppose that \(\beta_S=\beta\), \(\gamma_S=\gamma\) and we keep the value of \(\rho\). Using these parameters, and assuming that there is only one asymptomatic individual in the beginning of the simulation, we estimate the parameters \(\beta_A\), \(\gamma_A\) and \(p\) in order to minimize the squared error  
\begin{equation}
\begin{array}{cc}
\min_{\beta_A, \gamma_A, p} & \frac{1}{2} \left( \sum_{t}{ f \left( [(I_t^{cum} - D_t) - (\hat{I}_{S,t}+\hat{R}_{S,t})]^2 \right) + f\left([D_t - \hat{D}_t]^2\right)} \right)
\end{array},
\label{eq:SIRASD_model_params_calculation}
\end{equation}
\noindent where $I_t^{cum}$ and $D_t$ are real data provided by the Ministry of Health of Brazil, the cumulative number of infected individuals and deaths, and $\hat{I}_{S,t}$, $\hat{R}_{S,t}$ and $\hat{D}_t$ are estimated values of the symptomatic infected and recovered individuals, and deaths.

Table~\ref{tab:epidemiologicalParameters} presents the epidemiological parameters of our model\footnote{We also show other the epidemiological parameters obtained from other simulations in Table~\ref{tab:epidemiologicalParametersByRandomSeed} in the Section ``Methods''.} and some reference values. Some of the lines of this table deserve remarks. First, the basic reproductive number \(R_0\) in both models are comparable to the values for China and Italy. Second, the death rate \(\rho\) is very close to the values disclosed by the Brazilian Ministry of Health and the average of international values. We point out that our estimation of the death rate uses data that presumes there are places in hospitals to treat patients with severe infections, that is the situation that is present in the data now. Depending on the government policy, we do not know whether this is true or not at the peak of infection. Third, the proportion of symptomatic individuals \(p\) is smaller than the international reference due to the Brazilian Ministry of Health policy ``only test if you have strong symptoms''. In fact, the same problem of underdiagnosis also seems to have happened in the early epidemics in China \citep{Nishiura2020b}. 

In the last step of the estimation procedure, in order to estimate the parameter \(\psi\), we keep all model parameters as previously estimated and we also minimize the mean squared error using loss functions similar to the ones defined in Eqs. (\ref{eq:SIR_model_params_calculation}) and (\ref{eq:SIRASD_model_params_calculation}), depending on the case, in the period after March 23, 2020. Furthermore, in order to evaluate the effectiveness of the social distancing policy, we estimate a new value of \(\psi\) for each new point of the time series as shown in Table \ref{tab:psiOverTime}, where the column 2 shows the estimations of \(\psi\) for the SIRD model and column 4 shows estimations of the same parameter for the SIRASD model. Although there is a small gap between the  values of \(\psi\) for different models (SIRD or SIRASD), both columns suggest that the social distance factor \(\psi\) is going down, meaning that more people are joining the government policy. According to the models, the transmission rate is reduced to approximately 62\% of its original value. Table \ref{tab:psiOverTime} also presents the effective reproductive number derived from the impact of \(\psi\) on the transmission factors.

\subsection*{Forecasts}

\begin{figure}
    \centering

\begin{tabular}{c}
\includegraphics[width=80mm]{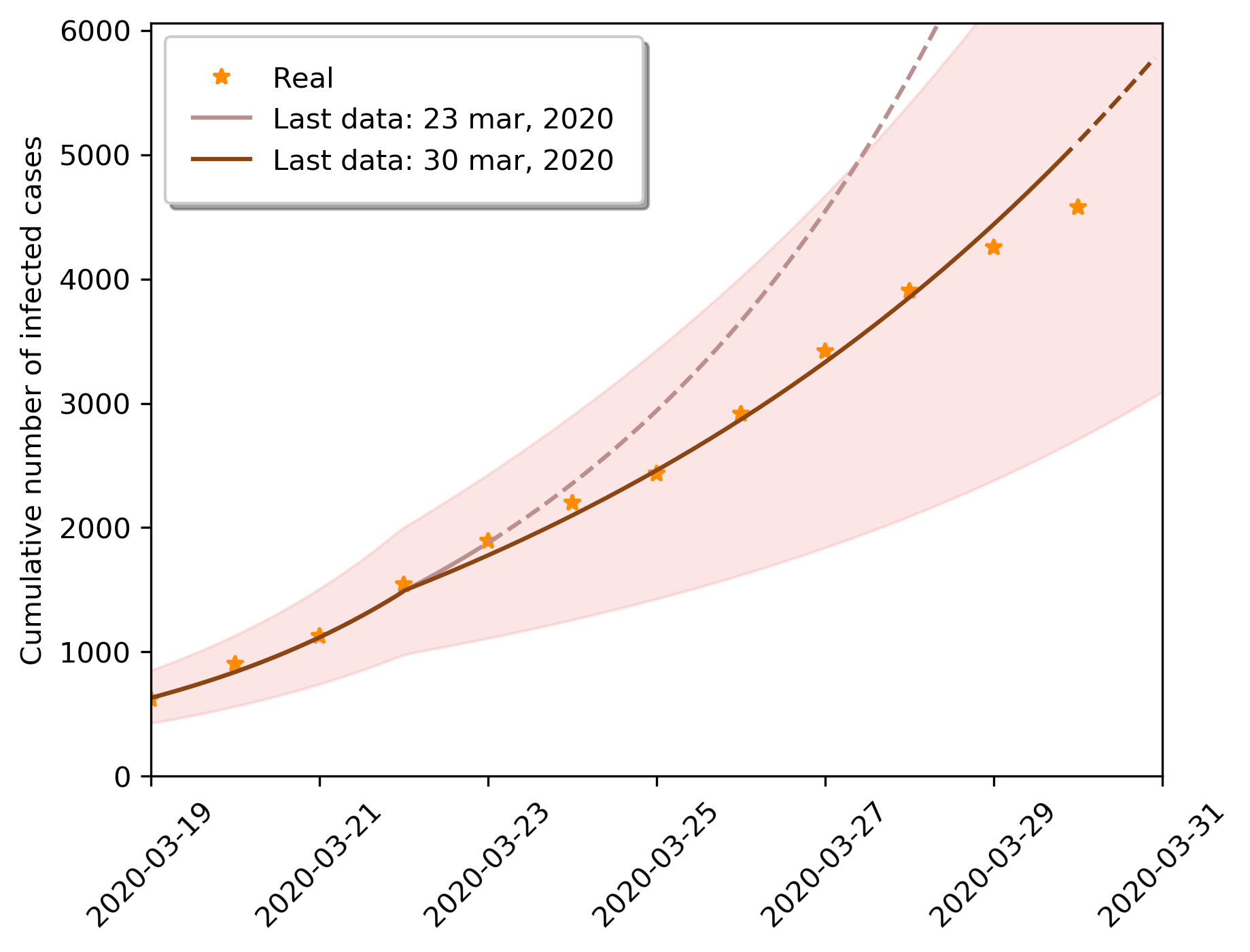}\\
\end{tabular}
\caption{Short term forecast of the SIRD model taking into account government social distance measures.  The solid line corresponds to the last date which the model was estimated, and the dashed line are model predictions. We show the evolution of the cumulative number of infected with 95\% confidence interval. We represent the real data as points.}
\label{fig:shortTermForecastSIR}

\end{figure}

\begin{figure}
    \centering

\begin{tabular}{c}
\includegraphics[width=80mm]{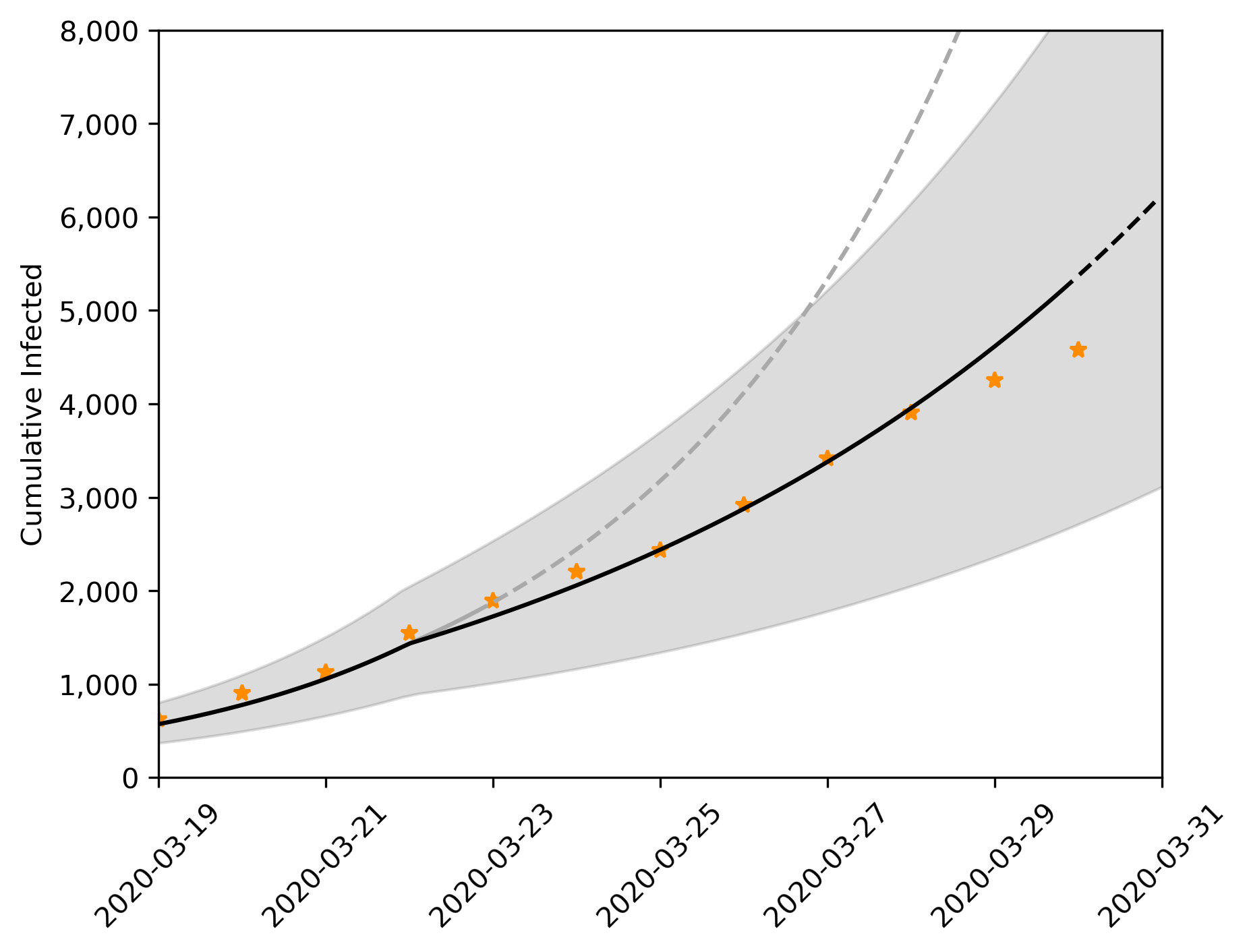}\\
\end{tabular}           
\caption{Short term forecast of the SIRASD model taking into account government social distance measures. The solid line corresponds to the last date which the model was estimated, and the dashed line are model predictions. We show the evolution of the infected (assymptomatic, symptomatic and both) with 95\% confidence interval. We represent the real data as points.}
\label{fig:shortTermForecastSIRASD}

\end{figure}

Figures \ref{fig:shortTermForecastSIR} and \ref{fig:shortTermForecastSIRASD} present respectively the short-term forcasts of the SIRD and the SIRASD models, where the models incorporate the $\psi$ factor in order to rescale the transmission factors ($\beta$, $\beta_A$ and $\beta_S$) in the scenario with the social distancing policy imposed by the government. Note that Figure \ref{fig:shortTermForecastSIRASD} explicitly shows the proportion of unknown asymptomatic individuals that when added to the symptomatic individuals skew the total value of infected individuals upwards.

\begin{figure}
    \centering

\begin{tabular}{c}
          \includegraphics[width=80mm]{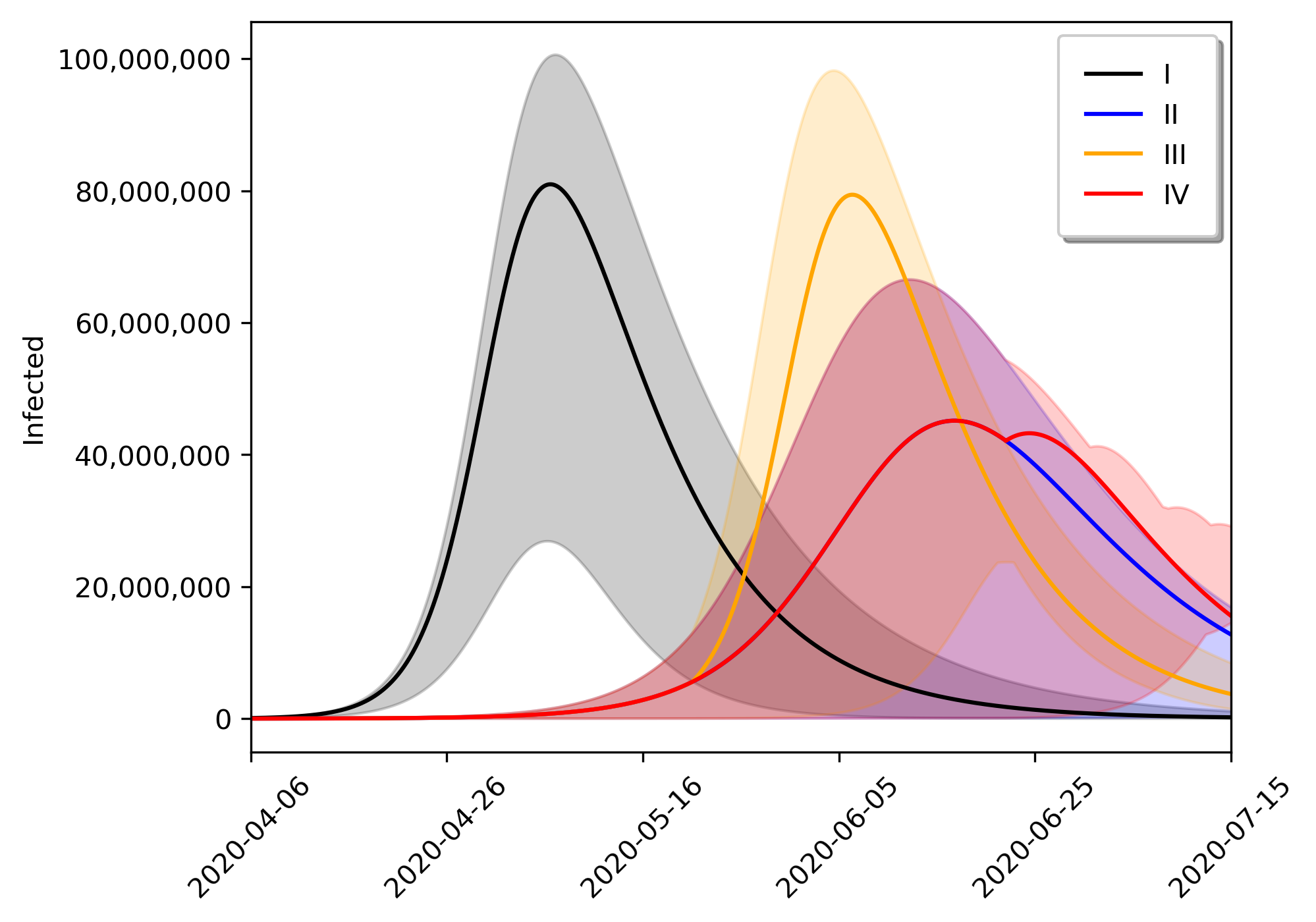}\\
\end{tabular}
\caption{Long term forecasts of number of infected for different scenarios using the SIRD model. Black, blue, yellow and red lines represent scenarios I to IV, respectively.}
\label{fig:longTermSIR}

\end{figure}

\begin{figure}
    \centering

\begin{tabular}{c}
          \includegraphics[width=80mm]{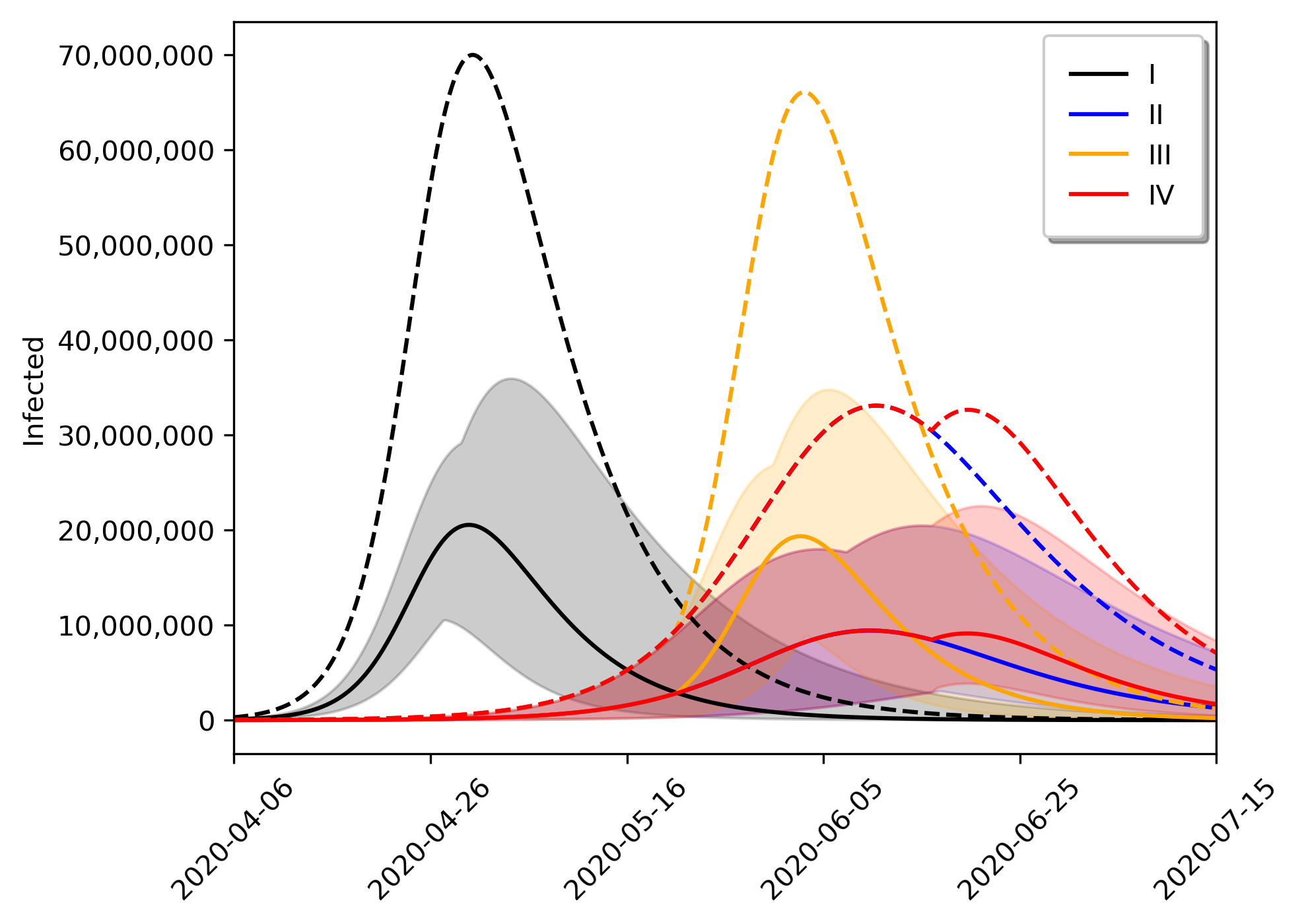}\\
\end{tabular}           
\caption{Long term forecasts of number of infected for different scenarios using the SIRASD model. Black, blue, yellow and red lines represent scenarios I to IV, respectively. While solid lines represent the symptomatic infected individuals, dashed lines represent total infected individuals.
}
\label{fig:longTermSIRASD}

\end{figure}

We also use the SIRD and SIRASD models to provide long term forecasts of the evolution of the COVID-19 pandemic in Brazil depending on the social distancing policy considered. While Figure \ref{fig:longTermSIR} shows the forecasts for the SIRD model, Figure \ref{fig:longTermSIRASD} shows the forecasts for the SIRASD model. In particular, we may note that while the SIRASD model predicts that the number of infected is higher than the estimates of the SIRD model, it also predicts a lower peak for the infected with symptoms, which are the ones that could require medical attention.

We explore four cenarios: (I) no measures of social distancing policy (black line); (II) current social distancing policy imposed by the government for an indefinite time (blue line); (III) 2-month social distancing policy imposed by the government (yellow line); and (IV) optimum limited time social distancing policy imposed by the government, so that the second infection peak is not greater than cenario II (red line). Scenario III suggests that policies based on short-term social distancing policy are not enough to constrain the evolution of the pandemic, that is, if social distancing policy measurements are released before the optimal time, a second peak should be experienced. The peaks and dates in which they occur are detailed in Table \ref{tab:peaks}. In the case of Scenario IV, the last day of the social distancing policy is June 22, 2020 for the SIRD model and June 16, 2020 for the SIRASD model.

\begin{figure}
    \centering
\begin{tabular}{cc}
\includegraphics[width=80mm]{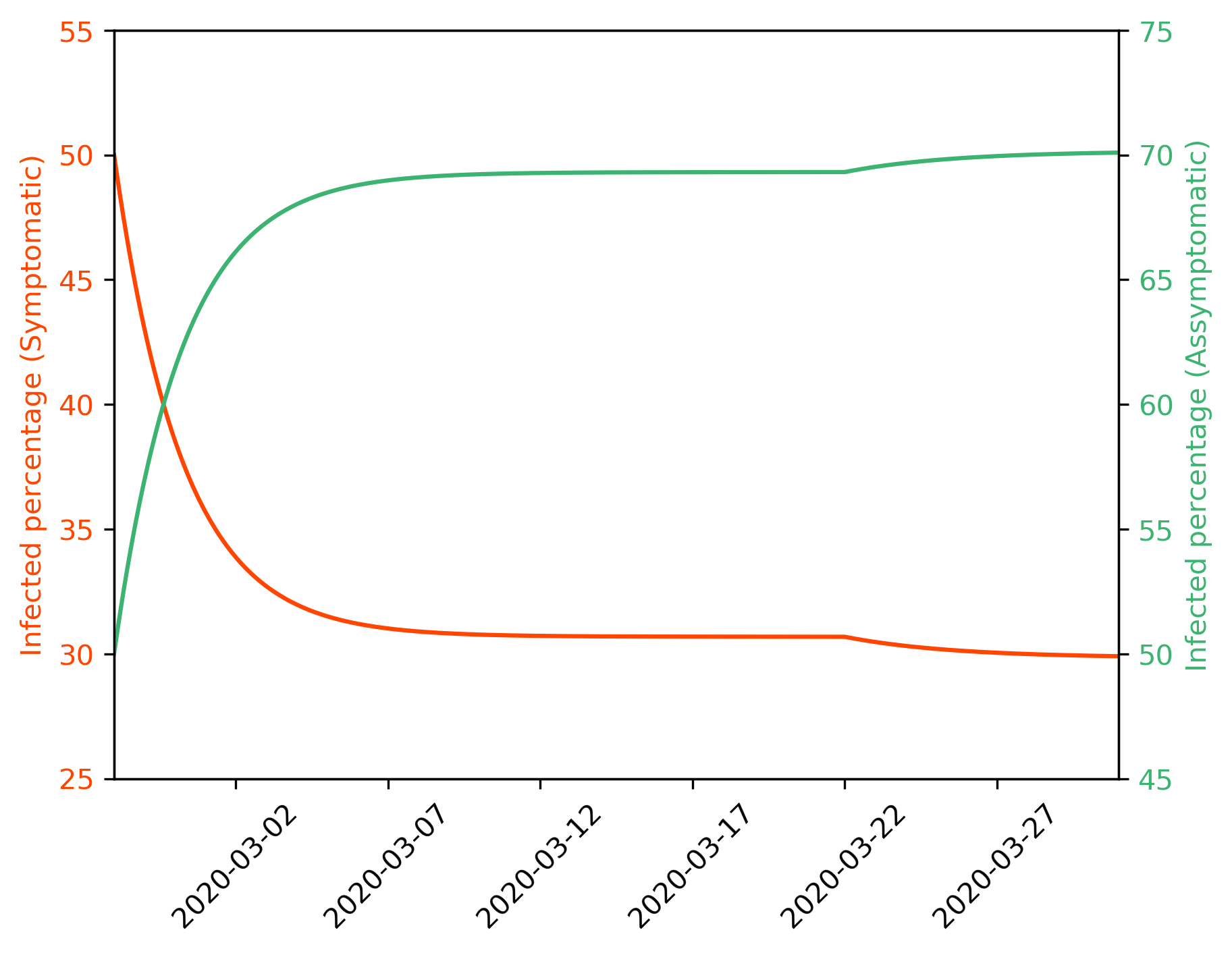} & \includegraphics[width=80mm]{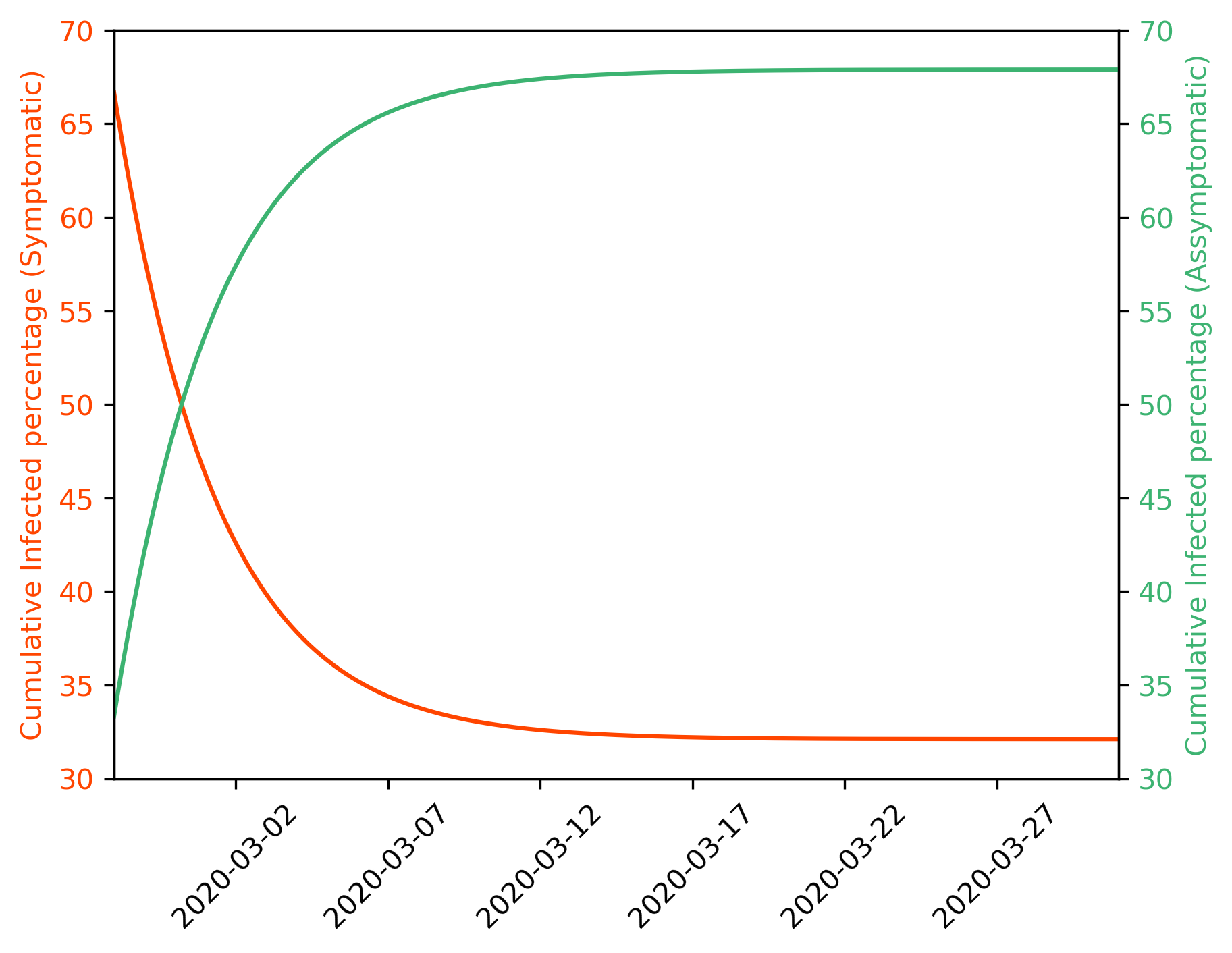} \\
(a) & (b) \\
\end{tabular}
\caption{Proportions of asymptomatic and symptomatic over time using $I_{A,0} = 1$. We show the instant proportion of infected (left) and the cumulative number of infected (right). Approximately 70\% are asymptomatic in March 30, 2020, which corresponds to 68\% cumulatively or $(1-p)$.}
\label{fig:symAsym}

\end{figure}

In addition to Figure \ref{fig:longTermSIRASD}, we also present the evolution of the proportion of asymptomatic and symptomatic in Figure \ref{fig:symAsym}. In this figure, we show the instant proportion [$I_{S,t}/(I_{A,t}+I_{S,t})$ for the symptomatic and $I_{A,t}/(I_{A,t}+I_{S,t})$ for the asymptomatic] and the cumulative proportion as well. Note that the proportion of individuals who develop symptoms, $p$ in Eq. (\ref{eq:IAISwithP}), alters the transmission rate, so it also affects the evolution of the number of asymptomatic and symptomatic individuals over time. So this plot estimates the evolution of this proportion. The last column of the last line of Table \ref{tab:epidemiologicalParameters} shows that the proportion of asymptomatic may vary from 29\% to 37\%, but this value is not fixed and evolves over time \citep{Mizumoto2020}. Our estimates suggest that the proportion of cumulative asymptomatic is approximately 68\% in March 30, 2020, which converged to $1-p$ (with $p$ given in Table~\ref{tab:epidemiologicalParameters}); that may account for some individuals with mild symptoms that were not tested.

\begin{table}
\hspace*{-1cm}
\centering
 \begin{tabular}{M{2.5cm}@{\extracolsep{4pt}}M{1.8cm}M{2cm}@{\extracolsep{4pt}}M{1.8cm}M{2cm}@{\extracolsep{4pt}}M{1.8cm}M{2cm}}
 ~ & \multicolumn{2}{c}{\textbf{SIRD}} & \multicolumn{4}{c}{\textbf{SIRASD}} \\ 
 \cline{2-3} \cline{4-7}
 ~ & \multicolumn{2}{c}{Infected ($I$)} & \multicolumn{2}{c}{Infected ($I_A+I_S$)} & \multicolumn{2}{c}{Symptomatic ($I_S$)} \\ \cdashline{2-3} \cdashline{4-5} \cdashline{6-7}
 \textbf{Scenario} & \textbf{Peak (\%)} & \textbf{Date} & \textbf{Peak (\%)} & \textbf{Date} & \textbf{Peak (\%)} & \textbf{Date} \\ \cline{1-1} \cline{2-3} \cline{4-7}
 I (Black)    & 38.5 & May 7   & 33.3 & April 30 & 9.7 & April 30 \\ 
 II (Blue)    & 21.4 & June 17 & 15.7 & June 10  & 4.4 & June 10 \\ 
 III (Orange) & 37.7 & June 6  & 31.4 & June 3   & 9.2 & June 3 \\ 
 IV (Red)     & 21.4 & June 17 & 15.7 & June 10  & 4.4 & June 10 \\ 
 \end{tabular}
 \caption{Peaks in each scenario and the dates of occurrence.}
\label{tab:peaks} 
\end{table}

\begin{figure}
    \centering
\begin{tabular}{c}
          \includegraphics[width=80mm]{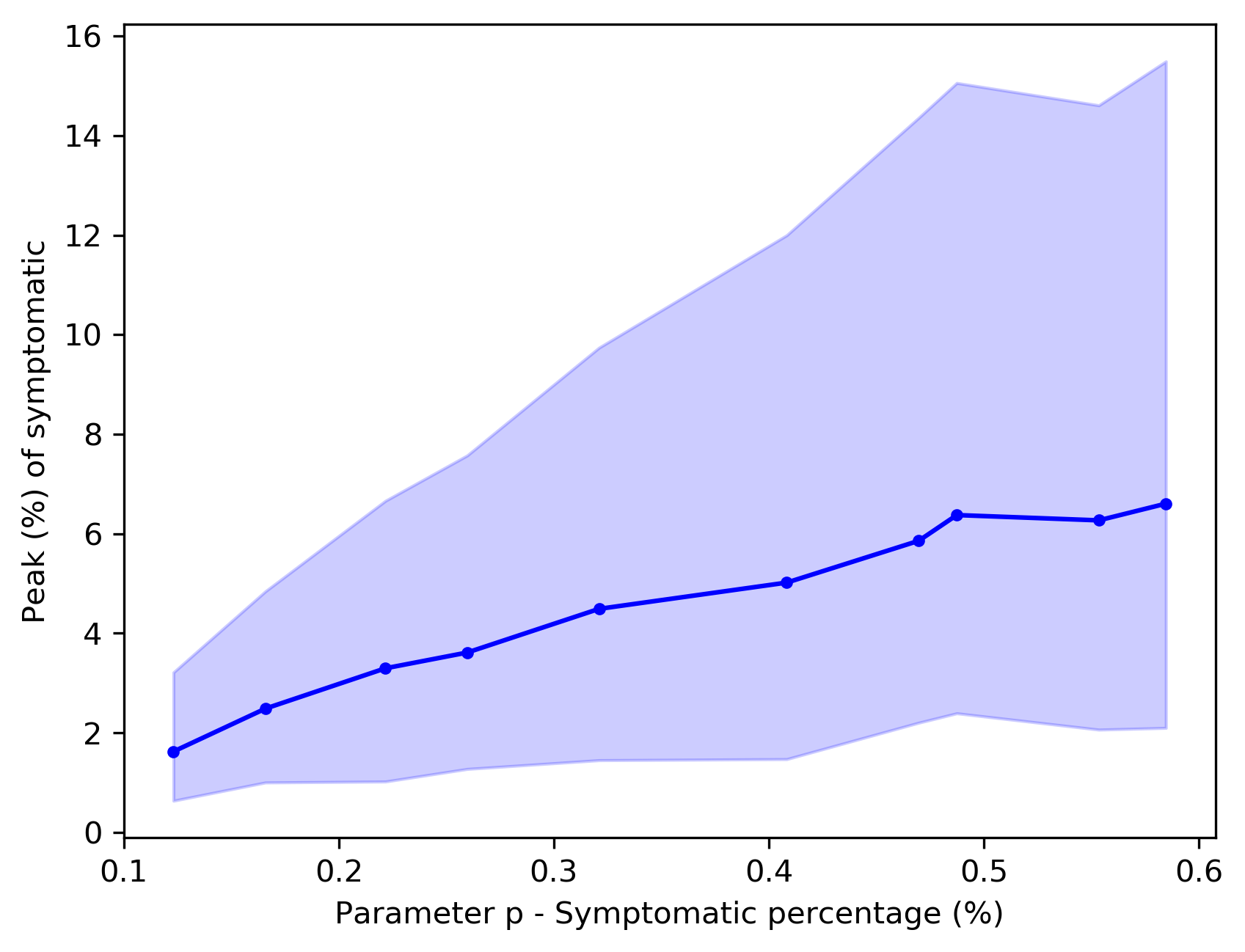}\\
\end{tabular}
\caption{The effect of symptomatic percentage (parameter $p$) in the proportion of symptomatic in the peak.}
\label{fig:longTermSIRASD_other_x0}

\end{figure}

Finally, it is worth considering that the SIRASD differential equations, presented in Eq. (\ref{eq:IAISwithP}), need an initial condition for the number of asymptomatic individuals. If we find the parameters values $(\beta_A, \gamma_A, p)$ by solving the optimization problem of Eq. (\ref{eq:SIRASD_model_params_calculation}) using different conditions, we get different results, that is, different peak values for the symptomatic individuals. If the proportion of asymptomatic individuals is larger, then this may be good news since it may represent less pressure for the health care system. But since we do not have enough tests to map the whole population, we need to work with hypotheses. 
Figure \ref{fig:longTermSIRASD_other_x0} shows the effect of different initial conditions in the symptomatic percentage and the peak value of symptomatic, that is, we vary the initial conditions, evaluate the symptomatic proportion (parameter $p$ in the SIRASD model), then calculate the peak value of symptomatic infected. So if we assume that the number of asymptomatic (symptomatic) individuals in data is larger (smaller) today, the number of asymptomatic (symptomatic) individuals will also be larger (smaller) in the time of the peak, leading to a smaller peak for the symptomatic. 

\section*{Discussion}

We use the Brazilian recent data from February 25, 2020 to March 30, 2020  to model and forecast the evolution of the COVID-19 pandemic in Brazil.

We estimate two variations of the SIR model using historical data and we find parameters that are in accordance with the international literature.  We also introduce a factor $\psi$ to account for the effect of the government social distancing measures. Our methodology is able to estimate the asymptomatic individuals, that may not be entirely present in data. Since the Brazilian government does not have enough tests for mass testing, this measure may provide some additional information. In fact, we show the relevance of the number of asymptomatic individuals, since the larger the number of asymptomatic individuals, the smaller the number hospital beds needed. The ``stay at home'' and ``only test if you have strong symptoms'' policies present contradictory effects in the disease control. While they avoid an increase in the number of infected people and the use of extra resources with people that present only mild symptoms, they reduce the amount of information about the real number of infected individuals. In particular, it explains the low value of the parameter that measures the proportion of individuals who present symptoms, since we count many individuals with mild symptoms as asymptomatic. 

While our short-term forecasts are in great accordance with the data, our long-term forecasts may help us to discuss different types of social distancing policies. We also show that the social distancing policy imposed by the government is able to flatten the pattern of contamination provided by the COVID-19, but short-term policies are only able to shift the peak of infection into the future keeping the value of the peak in almost the same value. Furthermore, we define the idea of the optimal social distancing policy as the finite social distancing policy that the second peak that happens after stopping the policy is not larger than the first. Based on this definition, we provide an estimate of the optimal date to end the social distancing policy.

An important discussion is about the effectiveness of vertical containment policies, where only people at risk follow social distance policies. In these kinds of policies, the two fractions of the population, the one at risk and the other one, present very different behaviors. First, the dynamics of the population at risk behaves similarly to the case with social distancing measures, but with a higher death rate. Second, the dynamics of the population that is not at risk behaves similarly to the case without social distance measures but with a low death rate. Third, since the fraction of the population that is at risk is much smaller than the rest of the population, the number of infected of the total population behaves similarly to the case without control. In fact, the policy's effectiveness is not in reducing the number of infected, but in reducing the number of deaths by confining individuals at risk. It is worth mentioning that the effectiveness of these vertical containment polices depends strongly on the ability to separate the individuals at high risk from the individuals at low risk and on the number of vacancies in hospitals to treat the disease. We may extend our model to explore these type of scenarios and we leave for future work.

Finally, another interesting research path is to evaluate the economic side effects of pandemic control  \citep{NBERw26882,Gormsen2020} and to propose measures to minimize these impacts \citep{Hone2019}.  

\section*{Methods}

\begin{table}
\centering
 \begin{tabular}{M{2.5cm}M{1.5cm}M{3.0cm}}
 \textbf{Model} & \textbf{Parameter} & \textbf{Interval of initial conditions} \\\hline
 SIRD& $\rho$ & [0.01, 0.1]\\\hline
 SIRD& $\beta$ & [1/10, 1/0.5] \\\hline
 SIRD& $\gamma$ & [1/14, 1/2]  \\\hline
 SIRASD& $\beta_S$ & $\{\beta\}$ \\\hline
 SIRASD& $\gamma_S$ & $\{\gamma\}$  \\\hline 
 SIRASD&$\beta_A$ & [1/10, $\beta_S$] \\\hline
 SIRASD&$\gamma_A$ & [1/14, 1/2] \\\hline
 Both Models &$\psi$ & [0, 1] \\\hline
 \end{tabular} 

 \caption{Parameters estimation region.}
\label{tab:parametersEstimationRegion} 
\end{table}

\subsection*{The solution of  the systems of  differential equations}

We find the numerical solutions of Eqs. (\ref{eq:SIRDead}) and (\ref{eq:IAISwithP}) through integration using the explicit Runge-Kutta method of order 5(4)  \citep{Dormand1980}. While this method controls the error assuming accuracy of the fourth-order, it uses a fifth-order accurate formula to take the steps. We use the implementation ``solve\_ivp'' of the scipy Python's library. 

The solution of the systems of differential equations depends on the definition of initial conditions. We use \(N_0=210147125\), that is the Brazilian population according to Brazilian Institute of Geography and Statistics (IBGE) which is the agency responsible for official collection of statistical, geographic, cartographic, geodetic and environmental information in Brazil, for both models. For the case of the SIRD model, we use $S_0=N_0-1$ and $I_0=1$. For the case, SIRASD model, we use 
$I_{S0} = 1$ and $S0 = N_0 - I_{A0} - I_{S0}$. We use $I_{A0}=1$ in all simulations of the paper but the simulations presented in Figure \ref{fig:longTermSIRASD_other_x0}, since we want to learn about the effect of $I_{A0}$ in the proportion of symptomatic and asymptomatic
individuals in the peak date.

\subsection*{The estimation procedure}

Our estimation procedure requires simultaneous integration of the differential equations (SIRD or SIRASD model depending on the case) and minimization of the loss functions [(\ref{eq:SIR_model_params_calculation}) or (\ref{eq:SIRASD_model_params_calculation})] depending on the case for each time \(t\). We minimize the loss functions using the method ``optimize.least\_squares'' also from the scipy Python's library \citep{2020SciPy-NMeth}  using the cauchy loss with scaling parameter $C=2$ \citep{Mayorov2015,Triggs1999}. 
To minimize the impact of the initial point assumption and data incompleteness, we repeat the estimation procedure 100 times using random initial conditions, but we discarded estimations which did not converge. Since this is a difficult nonlinear problem we bound the parameters estimation region. In particular, we use the bounds presented in Table \ref{tab:parametersEstimationRegion}. To be clear, the fact that $\beta_S=\beta$ and \(\gamma_S=\gamma$ is a consequence of our hierarchical estimation procedure previously described in the ``Results'' section. Furthermore, $\beta_A\in [0,\beta_S]$ means that \(\beta_A\le\beta_S\) \citep{Robinson2013}, since the asymptomatic individuals do not have symptoms that may help the spread of the infection.

Finally it is worth mentioning that this estimation procedure is sensitive to the random seed used by the algorithm as an initial condition. In particular, depending on this seed, we have found different epidemiological parameters in different simulations of the SIRD model, as presented in Table \ref{tab:epidemiologicalParametersByRandomSeed}. We have chosen the simulation results that provided the closest value of the median of the parameter $\gamma$, which is the one that used the random seed 7. We emphasize that although any of the presented epidemiological parameters could be a possible estimation and we could use them in the main part of this paper, this choice does change the qualitative analysis and the conclusions of our paper.

\begin{table}[hb]
\centering
\begin{tabular}{cccc}
 \textbf{Random seed} & $\boldsymbol{\beta}$ & $\boldsymbol{\gamma}$ & $\boldsymbol{\rho}$ \\ \hline
7 & 0.441717 & 0.150876 & 0.0292182 \\
511 & 0.432978 & 0.141411 & 0.0301936 \\
1024 & 0.428903 & 0.136487 & 0.032915 \\
90787 & 0.465757 & 0.176799 & 0.0273557 \\
407850 & 0.449369 & 0.159107 & 0.029013 \\
1905090 & 0.46357 & 0.174802 & 0.0277514 \\
\end{tabular}
\caption{\label{tab:epidemiologicalParametersByRandomSeed} Random seeds used in simulations and their respective estimated values of the epidemiological parameters.}
\end{table}


\subsection*{The long term forecasts}

The long term forecasts use the estimations presented in Table \ref{tab:epidemiologicalParameters} and the integration of the systems of differential equations as described in the beginning of this section. We build the 95\% confidence intervals of these curves randomizing the values of the parameters in the 95\% confidence intervals presented in Table \ref{tab:epidemiologicalParameters}.

\section*{Acknowledgment}

The second author is indebted to CNPQ for partial financial support under grant 302629/2019-0.




\end{document}